# Superconducting Power Generation

Mario Rabinowitz
Armor Research, lrainbow@stanford.edu
715 Lakemead Way, Redwood City, CA 94062-3922

**Abstract**

The superconducting ac generator has the greatest potential for large-scale commercial application of superconductivity that can benefit the public. Electric power is a vital ingredient of modern society, and generation may be considered to be the vital ingredient of a power system. This articles gives background, and an insight into the physics and engineering of superconducting power generation.

**Introduction**

With respect to the electric power industry, the superconducting ac generator has the greatest potential for large-scale commercial application of superconductivity. Such a machine should be able to convert mechanical energy to electric energy more efficiently and with greater economy of weight and volume than any other method. These advantages can be accrued at a scale of 1200 MVA output, with the added potential of operation at transmission line voltage, and greater system stability. In the past, a great deal of R & D was done in this area, but the present industry trend to smaller machines has decreased this effort. Though the advantages diminish at the much smaller scale of 250 MVA. such machines still offer interesting possibilities.

The overall reliability of five 250 MVA generators can be expected to be greater than of one 1200 MVA generator. The loss of one 250 MVA machine clearly has much less impact than the loss of one 1200 MVA machine. The advantage of reliability may outweigh the advantage of economy of scale to which large conventional machines are limited. Nevertheless, I believe that the advantages (as we shall see) of a 1200 MVA superconducting generators can be so much greater and so manifest, that at least for replacement purposes they would be seriously considered by utilities. "Build it and they will use it" applies here since enlightened utilities do take advantage of products that are clearly demonstrated to be superior..

Although quite different in their functions, generators and electric motors are essentially the same machine operated in inverse modes. A generator converts mechanical energy to electrical, and when the process is reversed it operates as a motor. There is a companion article in this issue on superconducting motors. Superconducting synchronous generators with a superconducting adjustable field rotor keep power losses to a minimum since the field in the stator is phase-locked in synchronism with the rotating rotor field. The high magnetic flux density produced by a superconducting rotor field winding permits a great reduction in the amount of iron required in both the rotor and stator. As we shall see, this reduction introduces degrees of freedom not previously possible in generator or motor design.

This paper is written to help us better perceive the technological potential of new developments. Let us start by gaining a simple understanding of the nature of superconductivity with both its promise and limitations.

**Nature of Superconducting Power Losses**

A superconductor at low or high temperature may be thought of as composed of two variable interpenetrating fluids. One fluid consists of normal conducting electrons. The other fluid consists of superconducting electron pairs. As the transition temperature is approached, the electrons all become normal. As the temperature approaches 0 K (absolute zero), all the free electrons become superconducting. This is like a parallel circuit with two branches. The superconducting electrons are represented by the branch with pure inductance and no resistance. The branch that represents the normal electrons has both inductance and resistance.

For dc, the current divides inversely proportional to the resistance of each branch. Thus the current goes solely through the superconducting branch without power loss, since there is no current in the normal (resistive) branch. For ac, the current divides inversely proportional to the impedance of each branch. Since the superconducting branch has inductive reactance, current flows in both branches with power loss in the resistive branch. This fundamental loss is always present.

Fortunately the fundamental loss is relatively small and generally negligible. However, there are other losses such as hysteresis and fluxoid motion power losses which may not be negligible. The power loss is relatively small in the dc rotor, as it is shielded from time varying fields from the ac stator. As explained in the introductory article in this issue, there are also maximum limits on current density and magnetic field even for dc. The large time varying fields at the stator result in hysteretic ac superconducting power losses that are presently too large to allow replacement of the copper in the stator coils with superconductor. However, this may change as ac losses are reduced, operating temperature is increased, and actual refrigeration efficiency is improved.

In superconducting technology applications, the devil lies in a myriad of details that require attention. For example, Parker and Krefta [1] point out that an important factor in the design of a superconducting rotor as well as any superconducting magnet is the interface electrical resistance between normal metals, or between a normal metal and a superconductor. For example, the interface resistance between the Al stabilizer and the Cu matrix of a filamentary superconductor as used in a rotor is very important to achieve stability. Under time varying fields as when the rotor field is ramping up or down, the interface resistance between the copper matrix and the superconducting filaments in magnet cables and ac transmission lines is an important factor in determining matrix losses. Parker and Krefta's analysis shows that there is an intrinsic magneto-resistance of cryogenic joints due to Hall currents.

**Historical Perspective**

The earliest work on superconducting generators was done in the U. S. in 1964 - 1966 in which an alternator was successfully demonstrated by Stekly, Woodson et al [2] using a fixed superconducting field coil, and a rotating normal armature with slip ring connections. The first rotating superconductor field coil was shown to be practical in 1971 by an MIT group. [3] Within a few years, R & D on similar superconducting synchronous machines was being conducted around the world.

In 1977, as manager in charge of superconducting projects at EPRI, I initiated two competing superconducting generator programs at Westinghouse and General Electric. Two objectives of these programs were to develop a detailed design of a 300 MVA superconducting generator, and a less detailed conceptual design of a 1200 MVA unit. I felt that it was important to look at a high enough power level to be meaningful for utility applications, and to have two independent and competitive projects. These were both successfully completed.

In 1994, the U.S. Department of Energy funded General Electric to design a 100 MVA HTS synchronous generator and build a full-scale HTS racetrack coil using BSCCO or TBCCO. The BSCCO coil achieved fields high enough for use in a generator with BSCCO supplied by Intermagnetics General Corp.

In 1988, the Ministry of International Trade and Industry (MITI) initiated the Super GM (Engineering Research Association for Superconductivity Generation Equipment and Materials) project, a collaboration of about a score of companies in Japan with a combined annual budget of about $50 million. One of their goals was to build and test three 70 MW superconducting ac generators using NbTi low temperature superconductor cooled by liquid helium. This was achieved in 1997 with three different rotors designed respectively for high stability, high critical current density, and low ac loss. [4] To my knowledge these are the largest superconducting generators built to date. These rotors have the capability of replacement of the NbTi with HTS material.

**Superconducting Generators**

There are at least four ways in which superconducting generators can substantially improve electric power systems:

#1.  They could greatly reduce both capital and operating costs.

#2.  They have the potential for high voltage operation thus eliminating the step up transformer; and thus they could also make dc economically competitive over considerably shorter distances than at present.

#3.  They would increase grid stability.

#4.  They could be used for var generation.

Let us see how these advantages can be accrued.

Voltage is induced and power is produced as the magnetic field lines of a rotating rotor periodically cut the stator windings.   A conventional generator with iron teeth has a magnetic field of $\approx$ 1 T (10,000 Gauss) at its inside surface.  This may be understood from the fact that the conductor slot width ~ tooth width, since iron saturates at $\approx$ 2 T. So a conventional generator operates between the limits of field saturation at the rotor and at the stator.  Superconducting rotor fields $\approx$ 5 T (50,000 Gauss) greatly exceed the saturation field of iron.  Thus iron can largely be eliminated in the stator (armature) and be replaced by copper for an increased density of ampere-turns induced by the rotating field.  This concomitantly reduces insulation requirements.

Both changes allow for better use of space and materials.  When grounded iron is removed from the stator windings, less insulation is needed between adjacent bars. Although full interphase insulation is still needed between end-winding groups. the reorganized windings and insulation (made possible by the absence of the stator's iron teeth) permit reduced insulation stress and, in turn, higher voltage operation.   This presents the potential for  the development of a high voltage superconducting generator which could entirely eliminate the step-up transformer to the transmission line.

Because a superconducting generator is a compact object, the refrigeration requirement is not a dominant factor as it is in a  superconducting transmission line. For machines bigger than 100 MVA,  even at low temperature operation, the refrigeration requirement is a small part of the overall losses even with the big gain in refrigeration efficiency at 77 K operation.  Nevertheless, higher temperature operation would be desirable from the aspect of increased reliability, simplicity, and tolerance to temperature rises.   Although operation at higher temperatures would be an advantage, the use of low temperature superconductors is quite adequate.

Despite being much improved relative to their inception, high temperature superconductors still have problems of greater brittleness and lower critical current density than do low temperature superconductors.  One possible way to get around these problems is with a stored field rotor in which the magnetic field is trapped in the superconductor with high fidelity to an original pattern field. [5]  As long as the superconductor is kept below the transition temperature for the given trapped field, the superconductor performs like a very high field permanent magnet.  A stored field rotor would also eliminate many of the problems associated with winding a superconducting coil, and losses due to wire motion (slip) in the coil.

Although the actual cost of an ac generator is relatively insignificant compared with the cost of a power plant, it may be considered to be the pivotal element of the plant.  Conventional generators are $\approx$ 99 % efficient.  An increase of 0.5% to a 99.5 % efficiency, made possible by superconducting generators, is equivalent to a much more impressive sounding 50 % reduction in losses.  But the power plant leveraging is much greater than this since the generator dominates not only the electrical aspects of the power station, but strongly impacts the overall capital costs of the plant as indicated by items #1 and #2.  The efficiency improvement is leveraged by a factor of $\approx$ 100 since the ratio of the capital cost/kW of a power plant are $\approx$  100 times higher than that of a generator.  Since superconducting generators are significantly smaller than conventional generators, there are substantial further savings in construction, financing, and operating costs.  Figure 1 and Table 1 make a comparison between conventional and possible superconducting 1200 MVA generators.  The reason that superconducting generators are significantly smaller than conventional ones of the same power output is because the power density of a generator is approximately proportional to the square of the magnetic excitation field, $\propto (5T/2T)^2 \approx 6$.

The potential for increased ac system stability afforded by a superconducting generator is also of great importance as noted in item #3. When a system (generator-grid-load) is perturbed electrically or mechanically, the resulting transient is oscillatory. If the oscillations damp out to the original or a new steady-state operating condition, the system is stable. Otherwise the system is unstable. In general, the lower the generator reactances, the greater the stability. The per-unit synchronous reactance of a superconducting generator is about 1/4 that of a conventional generator of similar rating. This results in an increased steady-state stability limit of the superconducting machine by as much as a factor of four when the transmission line reactance is relatively small.

A superconducting generator can also be operated as a synchronous condenser to adjust the reactance of a line, i.e. var generation as mentioned in item #4. Static var generators introduce harmonics into the system unless there is sufficient filtering. A superconducting synchronous condenser would not introduce harmonics at all.

**Natural and Artificial Limits**

Traditionally, both in generation and in transmission, electric utilities have pursued economies of scale with large power plants in increasing efficiencies and in reducing capital and operating costs. However, both natural and artificial constraints limit such expansion. A limit for conventional generators is about 1500 MVA. This is roughly a shipping limit for 4-pole machines, and somewhat of a natural limit for 2-pole machines since rotor rigidity is essentially proportional to $r^2/\ell^3$, where r is the rotor radius and $\ell$ is the rotor length. Even if r were not already limited by centrifugal force, it would be seriously limited by transport facilities. So a given rotor stiffness cannot be maintained simply by increasing r faster than $\ell$ is increased. As $\ell$ is increased to increase the power output, the natural frequency is decreased and with it a decrease in the first critical speed and an increase in the number of critical speeds. Thus more stringent and sophisticated balancing is required. For an end supported rotor, flexure and vibrations severely limit the length, and the rotor must continually be rotated to avoid sag.

Thus, machine size is ultimately limited by balancing or shipping capability. When these limits are reached, new technological breakthroughs such as the superconducting generator can result in new greater limits, or to new ways of doing business.

**Discovery and Commercialization**

Many advanced technology races for the electric power industry are being run all over the world in both basic research and for commercialization. In the field of superconductivity, experimental and theoretical efforts have been aimed at achieving a fundamental understanding of new high temperature superconductors (as well as the old low temperature) superconductors that will lead to improved materials in terms of higher transition temperatures, higher critical current densities, lower alternating power
losses. better mechanical properties, fabricability, etc.

In the race for fundamental knowledge, the United States has traditionally been at the forefront of scientific discoveries. However the race for commercialization has not always been so successful. For advanced technologies like high temperature superconductivity, we need to avoid the same loss of leadership that has befallen manufacturers of electric power industry apparatus (e.g., circuit breakers, transformers, and high power semiconductors); and on a broader scale the electronic and microelectronics industry (televisions, stereos, VCR's, etc.).

**Conclusion**

A well-designed superconducting generator can represent a beautiful optimization between competing and often conflicting electrical, economic, thermal, reliability, and mechanical requirements. [6] Basically thermal performance improves as the operating temperature is increased, and electrical performance is decreased. Electrical system stability and power density is increased at the expense of mechanical strength and vibration tolerance. The increased complexity of the advanced technology is not allowed to compromise reliability. Keeping capital costs down, compromises everything.

The emergence of a technology is a function of both its potential and of resources put into its development. In simple terms, the superconducting generator leads in potential because of its small surface to volume ratio i.e. it is relatively compact; and because the superconductor is operated in the dc mode. With respect to power loss, the superconducting ac transmission line is limited due to a very large superconducting surface to volume ratio, and hence large refrigeration losses, as well as relatively high power losses because it is operated in the ac mode.

Because the USA is a free and open society with respect to the sharing of scientific knowledge and to the exchange of ideas, some say we are more vulnerable in the marketplace. However, when we are united by common goals and coordinated implementation, the diverse interests of industry, federal laboratories, and universities can meet the nation's needs in scientific research as well as protecting our position in commercialization.

Even though the advantages of higher efficiency, reduced size and weight, lower capital cost, and greater stability may be realized in an ac superconducting generator, these advantages could easily be outweighed by one factor -- reliability. If it is less reliable than a conventional machine, even a few days more outage per year could well tilt the balance against it. It is a more complicated and more delicate machine than its rugged counterpart. In fact, part of its beauty stems from its achievement without recourse to brute force. It has the potential of being a solution where the past answers of more and larger are no longer adequate.

**Table 1. Comparison of Conventional and Possible Superconducting 1200 MVA Generators**

| Property | Superconducting Generator | Conventional Generator |
| --- | --- | --- |
| Phase-to-Phase Voltage | 26 to 500 kV | 26 kV |
| Line Current | 27 to 1.4 kA | 27 kV |
| Active Length | 2.5 to 3.5 m (8 to 11.5 ft) | 6 to 7 m (20 to 25 ft) |
| Total Length | 10 to 12 m (33 to 39 ft) | 17 to 20 m (55 to 65 ft) |
| Stator Outer Diameter | 2.6 m (8.5 ft) | 2.7 m (8.9 ft) |
| Rotor Diameter | 1 m (3 ft) | 1 m (3 ft) |
| Rotor Length | 4 m (13 ft) | 8 to 10 m (26 to 33 ft) |
| Field Exciter Power | 6 kW | 5000 kW |
| Generator Weight | 160 to 300 ton | 600 to 700 ton |
| Total Losses | 5 to 7 MW | 10 to 15 MW |
| Field Strength | 5 Tesla | 2 Tesla |